\documentclass[conference, a4paper, 10pt, fleqn]{IEEEtran}
\usepackage{cite}
\ifCLASSINFOpdf
  \usepackage[pdftex]{graphicx}
\else
  \usepackage[dvips]{graphicx}
\fi

\usepackage[cmex10]{amsmath}
\usepackage{amssymb}

\usepackage{array}
\usepackage{booktabs}
\usepackage{threeparttable}
\usepackage[update,prepend]{epstopdf}
\usepackage{multirow}

\usepackage[top=20mm, bottom=20mm, left=15mm, right=15mm]{geometry}
\begin{document}
\title{\bfseries \Large Prediction-based Adaptation (PRADA) Algorithm for Modulation and Coding}
\author{\IEEEauthorblockN{\large Shou-Pon Lin, Jhesyong Jiang, Wei-Ting Lin,  Ping-Cheng Yeh, Hsuan-Jung Su}
\IEEEauthorblockA{\normalsize Department of Electrical Engineering and\\
Graduate Institute of Communication Engineering\\
National Taiwan University\\
Taipei, Taiwan}}

\maketitle

\begin{abstract}
\boldmath
In this paper, we propose a novel adaptive modulation and coding (AMC) algorithm dedicated to reduce the feedback frequency of the channel state information (CSI). There have been already plenty of works on AMC so as to exploit the bandwidth more efficiently with the CSI feedback to the transmitter. However, in some occasions, frequent CSI  feedback is not favorable in these systems. This work considers finite-state Markov chain (FSMC) based channel prediction to alleviate the feedback while maximizing the overall throughput. We derive the close-form of the frame error rate (FER) based on channel prediction using limited CSI feedback. In addition, instead of switching settings according to the CSI, we also provide means to combine both CSI and FER as the switching parameter. Numerical results illustrate that the average throughput of the proposed algorithm has significant performance improvement over fixed modulation and coding while the CSI feedback being largely reduced.
\end{abstract}

\section{Introduction} \label{intro}
In recent years, the demand of wireless communication channel has grown tremendously. It may be not sufficient to support the large amount of data over a bandwidth-limited time-varying fading channel using fixed modulation and coding. The fixed modulation and coding schemes with power control are designed to maintain certain quality of service (QoS) when the channel condition is poor. Thus, fixed modulation and coding scheme does not fully exploit the time-varying nature of the channel. Adaptive modulation and coding (AMC), on the other hand, takes advantage of the time-varying characteristic to achieve high spectral efficiency while maintaining acceptable bit error rate (BER). The notion of adaptive transmission has been proposed striving to satisfy growing consumer demands \cite{goldsmithCap}. Moreover, some adaptive transmission techniques have been widely included in wireless standards such as IEEE 802.16\cite{mag802.16}, IEEE 802.22 \cite{mag802.22}, and LTE \cite{magLTE}.

Several pioneer works on AMC have been proposed in the literature. In \cite{goldsmithMQAM} and \cite{goldsmithCoded}, various rate and power adaptation is considered. Goldsmith et. al. \cite{goldsmithMQAM} proposes a variable-power variable-rate modulation scheme. The transmitter switches settings according to the channel state information (CSI) feedback by the receiver. The transmitter compares the received channel gain to a set of prescribed thresholds. The thresholds partition the channel gain into a set of regions corresponding to different adaptation settings, and the transmitter makes decision accordingly. Once a setting has been determined, the setting is unchanged for a transmission frame. The settings in \cite{giannakisARQ} and \cite{giannakisQueue} are restricted to variable-rate while the power is kept constant. In \cite{yun05}, a frame consists of a variable number of packets depending on the current channel state. The settings of \cite{giannakisARQ, giannakisQueue, yun05} are switched at the start of transmission of each frame. All of these works require current channel state information at the receiver,
thus a reliable feedback path is required for the CSI feedback. Moreover, the receiver is obliged to feedback the CSI every frame. 

However, in some circumstances, it is prohibitive to maintain a high feedback rate. For the systems where power is a concern, such as sensor networks \cite{marquess08sensor}, frequently CSI feedback is not available. Moreover, the research on adaptive transmission technique combined with multiple-input multiple-output (MIMO) or orthogonal frequency division multiplexing (OFDM) or both of them has already received more and more attention. The adaptive OFDM system \cite{duelhallen06LRPOFDM,choi08,cheeLump} involves optimizing the modulation level and transmission power over the entire frequency band, so the CSI is required for all subcarriers resulting in high feedback load. In MIMO systems \cite{love04magazine}, the multiple antennas lead to the increasing complexity of CSI. As a result, the CSI has higher dimension and is more complex in these systems than that in the single carrier case, and there have been efforts on the reduction of feedback load in these system.

Therefore, the further reduction of feedback overhead is desired. One approach is to reduce the frequency of CSI feedback from every frame to every $M$ frames. The adaptation period of the system is also reduced to every $M$ frames. However, in the previous works, adaptive transmission makes decision based on current channel state \cite{goldsmithMQAM, goldsmithCoded, giannakisARQ, giannakisQueue}, or the most probable channel state estimated by ACK/NAK signal \cite{karmokarPOMDP}. If the adaptation period is reduced to every $M$ frames, the number of the CSI received at the transmitter side decreases. Without the knowledge of channel variation between intervals of CSI feedback, the adaptive method is chosen only to maximize the throughput of the first frame of the $M$ frames instead of the throughput of all subsequent $M$ frames. Consequently, the performance of such system is poorer than the system with shorter adaptation period.

In order to maximize the overall throughput, it is required to take the channel state of subsequent $M-1$ frames into consideration. Thus, some predictions on the channel state of subsequent $M-1$ frames are to be made. Currently there are several channel prediction method such as linear prediction \cite{falahati03,oien04}, channel variation prediction based on finite state Markov chain (FSMC) \cite{chen99,cheeLump}, etc. However, in the work \cite{chen99,cheeLump}, only the channel state prediction of the next frame is considered. The settings are selected to maximize certain criteria of only the transmission of the next frame. Their results could not extend the adaptation period to $M$ frames. 

The previous works on channel prediction does not provide the analysis of BER or FER over long variation of channel. In this paper, a Markov chain-based channel prediction method is proposed. We are able to predict the channel variation of the following $M-1$ frames through the first frame and derive the close-form expression of the frame error probability of subsequent $M-1$ frames. Hence the channel variation of subsequent $M-1$ frames are taken into consideration, which enables the optimization the overall throughput of the $M$ frames. On the basis of channel prediction based on FSMC, we propose a novel adaptation algorithm: Prediction-based Adaptation (PRADA) Algorithm for modulation and coding. The proposed algorithm is able to extend the adaptation period from every frame to every $M$ frames while still maintaining good performance. The numerical result shows that with channel prediction, the throughput of system using PRADA increases significantly over the system that only maximizes the throughput of the first frame.

The systems proposed by the previous works \cite{goldsmithMQAM, goldsmithCoded, giannakisARQ, giannakisQueue, yun05} switch settings based on the received channel gain. There are also works proposing to utilize solely ACK and NAK feedback \cite{karmokarPOMDP}. Another approach does not assume the channel condition but simply use ACK and NAK as setting switching criterion \cite{falahatiHARQ}. The transmitter first selects the largest constellation which has the highest data rate and the highest error probability, and reduces the constellation size once receives retransmission request. The another approach uses ACK and NAK as an observation to estimate the underlying channel state \cite{vucetic91,karmokarPOMDP}. In \cite{karmokarPOMDP} the transmitter forms a belief state space using underlying FSMC model. The transmitter determines the most possible state according to the received ACKs or NAKs, and switches settings accordingly. On the other hand, in the algorithm proposed by Vucetic \cite{vucetic91}, the transmitter compares the block errors to the specified thresholds to estimate the current channel state. These thresholds are determined by minimizing the errorneous channel estimation. Both algorithms adopted the FEC codes of different rates correspond to the estimated channel state. 

In this paper, we develop another adaptation algorithm PRADA-B which combines both feedback-reduced CSI and ACK/NAK signal, while the precedent works use either CSI or ACK/NAK only. The transmitter compares the number of frame errors in $M$ frames to the thresholds  which are computed by the CSI feedback every $KM$ frames. The thresholds are determined with the view to maximize the overall throughput over consecutive $KM$ frames, instead of optimization on the current frame. Switching settings according to FER can further reduce the CSI feedback while maintaining acceptable throughput. The receiver feeds back the CSI to the transmitter at the beginning of every $KM$ frames in order to track the channel behavior.

The rest of this paper is organized as follows. The system and channel model is given in Section II. The adaptation algorithm of PRADA is given in Section III. We develop the mathematical basis of the algorithm, the frame error rate derivation and the threshold of setting switching in Section IV. The numerical results of the proposed algorithm are demonstrated in Section V. Finally, we draw concluding remarks in Section VI.

\section{System Model}

\subsection{System Overview}

\begin{figure}
\centering
\includegraphics[width=0.5\textwidth]{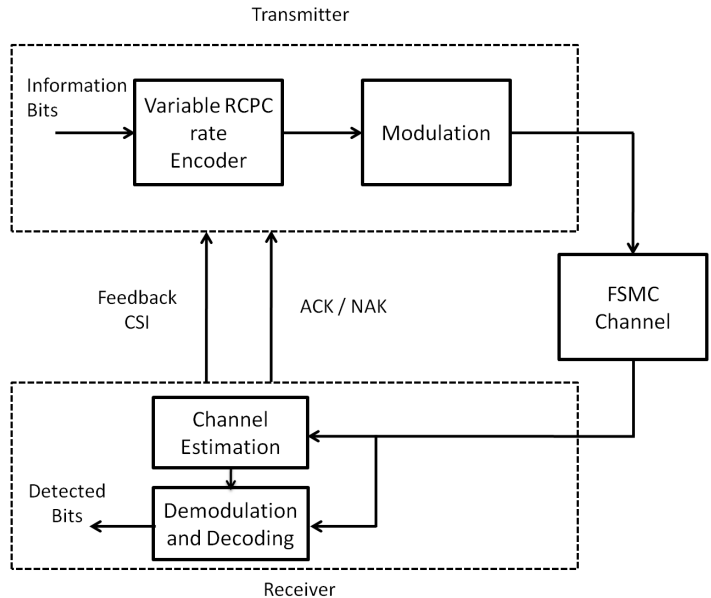}	
\caption{System model}
\label{fig_system_model}
\end{figure}

The system model considered in this work is given in Figure \ref{fig_system_model}. The information bits are encoded by the rate-compatible punctured convolutional (RCPC) code, modulated, then transmitted. The transmitter switches the modulation and the coding every setting adaptation period. The setting adaptation period is $M$ frame periods. 

In this model, $R$ settings are chosen from the combinations of  
modulations and RCPC codes. 
Denote $R$ settings as $\{s_1, s_2, \dots, s_R\}$, 
listed in the descending order of the spectral efficiency, which has the same meaning as listing the settings in the descending order of FER in this system. 
The FER is computed based on the same number of modulation symbols.
Each setting corresponds to a data rate, 
which combines the effect of the 
modulation and the coding. 
Denote the data rate of each setting as 
$\{k_1, k_2, \dots, k_R\}$.
Note that $k_1 \geq k_2 \geq \dots \geq k_R$ 
because the setting with higher data rate 
also has higher FER.

The receiver sends an ACK or a NAK to 
the transmitter when it receives a frame.
In addition, the receiver sends the CSI 
to the transmitter at certain frequency.
The transmitter switches the setting according
 to the CSI and the FER. 
The algorithms are described in Section \ref{adaptation}.

\subsection{Channel Model}\label{sec_channel_model}
FSMC is widely used to model Rayleigh 
fading channel which is a good approximation
 of channel with memory. And in the application 
of packet/frame transmission, first-order Markov
 chain's accuracy is supported by \cite{ccTan}. There are 
some studies on how to partition SNR into finite 
states of FSMC model. Wang and Moayeri 
\cite{wangFSMC} propose a method partitioning 
all FSMC states that have the same stationary probability.
 Zang and Kassam \cite{zhangFSMC} propose a 
method partitioning all FSMC states that have the 
same average time duration. Park and Hwang 
\cite{parkFSMC} propose a method partitioning 
FSMC states such that the non-adjacent transition 
probabilities are small enough. In this work, 
we adopt the channel model proposed in 
\cite{zhangFSMC} which is a better approximation 
and the average time duration is made large 
enough to meet our assumption that in each
 frame transmission the state will not change.

The channel SNR $\gamma$ is quantized into 
several non-overlapping states with boundary values 
$\mathbf{\Gamma} = \left[\Gamma_1 \Gamma_2 \dots \Gamma_{N+1}\right]$ 
with $\Gamma_{1}$ = 0 and $\Gamma_{N+1} = {\infty}$. 
The channel is said to be in state $w_i$ if 
$\gamma \in [\Gamma_i \Gamma_{i+1} )$. 
Thus we can obtain stationary probability $\pi_i$ 
of $w_i$ using the probability density function of Rayleigh fading channel.
\begin{equation}
\pi_i = \int_{\Gamma_i }^{\Gamma_{i+1}}\frac{1}{\gamma_0}\exp(-\frac{\gamma}{\gamma_0})d\gamma
\end{equation}
where $\gamma_0$ is the average SNR.

In this system, we assume constant time in each frame transition, denoted by $T_{p}$. We can obtain the transition probabilities
\begin{equation}
\begin{split}
P_{i,i+1} &= \frac{N(\Gamma_{i+1})T_p}{\pi_i} &i&=1,2,\dots,N-1\\
P_{i,i-1} &=\frac{N(\Gamma_{i})T_p}{\pi_i} &i& = 2,3,\dots,N
\end{split}
\end{equation}
where $N(\Gamma_{i})$ is the level crossing rate at SNR level $\Gamma_{i}$ either in the positive direction or in the negative direction, given by \cite{zhangFSMC}
\begin{equation}
N(\Gamma) = \sqrt{\frac{2\pi\Gamma}{\gamma_0}}f_m\exp(-\frac{\Gamma}{\gamma_0})
\end{equation}
where $f_m$ denotes the maximum Doppler frequency due to the user motion.
\subsection{FER Model} 

Given the channel model, we can analyze the performance of the system.
Let $P_{s_r}(\gamma)$ denote the FER of the channel SNR $\gamma$ when using setting $s_r$. The FER of the channel state $w_i$ when using setting $s_r$, which is denoted by $E^{w_i}_{s_r}$, can be derived as
\begin{equation}
E^{w_i}_{s_r}=\frac{\int_{\Gamma_i}^{\Gamma_{i+1}}P_{s_r}(\gamma)p_t(\gamma)d\gamma}{\int_{\Gamma_i}^{\Gamma_{i+1}}p_t(\gamma)d\gamma},
\end{equation}
where $p_t(\gamma)$ is the probability that channel SNR is $\gamma$.

In this work, the FER is obtained by computer simulation using RCPC codes. The modulation and the puncturing rate are corresponding to the setting $s_r$. However, the expression can also be computed using analytical expression or simulation-fitting expression of $P_{s_r}(\gamma)$.


\section{Algorithm Descriptions}\label{adaptation}

In this section, we describe the proposed algorithms, that is, PRADA-A and PRADA-B. PRADA-B considers the CSI feedback and the FER in previous transmissions. In order to choose the optimal setting, it requires a threshold table for FER comparison. Although the computation can be done offline, the complexity may be not allowed in some system. PRADA-A is also  proposed because it reduces the computational complexity of PRADA-B. PRADA-A considers only the CSI feedback and its computation is only the first part of PRADA-B'’s computation.

By simulation, we can get the FER of each setting in each state. The expected throughput of successive $M$ frames given the first channel state and the setting can be derived if the FER and the channel transition probability are known. It is used in both PRADA-A and PRADA-B. The derivation of the expected throughput is given in Section \ref{derive_throughput}. 

To deal with the doppler frequency variation, the algorithms use different FSMC channels. That is, we can generate several
FSMC channels for possible doppler frequency and use them in algorithm computation.

\subsection{PRADA-A Algorithm} \label{alg_prada_a}
\begin{figure}
\centering
\includegraphics[width=0.5\textwidth]{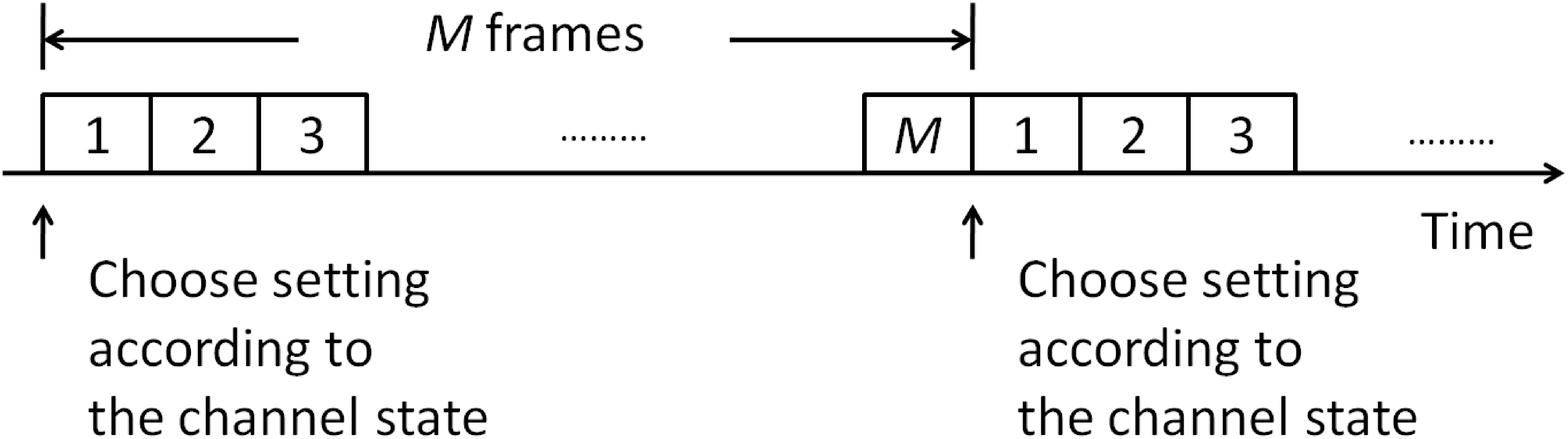}	
\caption{Time domain diagram of Prada-A algorithm}
\label{fig_prada_a}
\end{figure}

Figure \ref{fig_prada_a} illustrates the time domain diagram of PRADA-A. We call $M$ frame transmission duration a setting adaptation period. At the beginning of a setting adaptation period, the transmitter chooses the setting maximizing the throughput of the next $M$ frames according to the current channel state. Here we summarize the steps as follows:

\begin{itemize}
\item Step 1: For every setting, compute the expected throughput of the next $M$ frame transmissions 
for every possible beginning channel state. This can be done offline.
\item Step 2: At the beginning of a setting adaptation period, 
the transmitter observes the channel state and chooses the corresponding setting.
\item Step 3: Repeat 2.
\end{itemize}

\subsection {PRADA-B Algorithm} \label{alg_prada_b}
\begin{figure}
\centering
\includegraphics[width=0.5\textwidth]{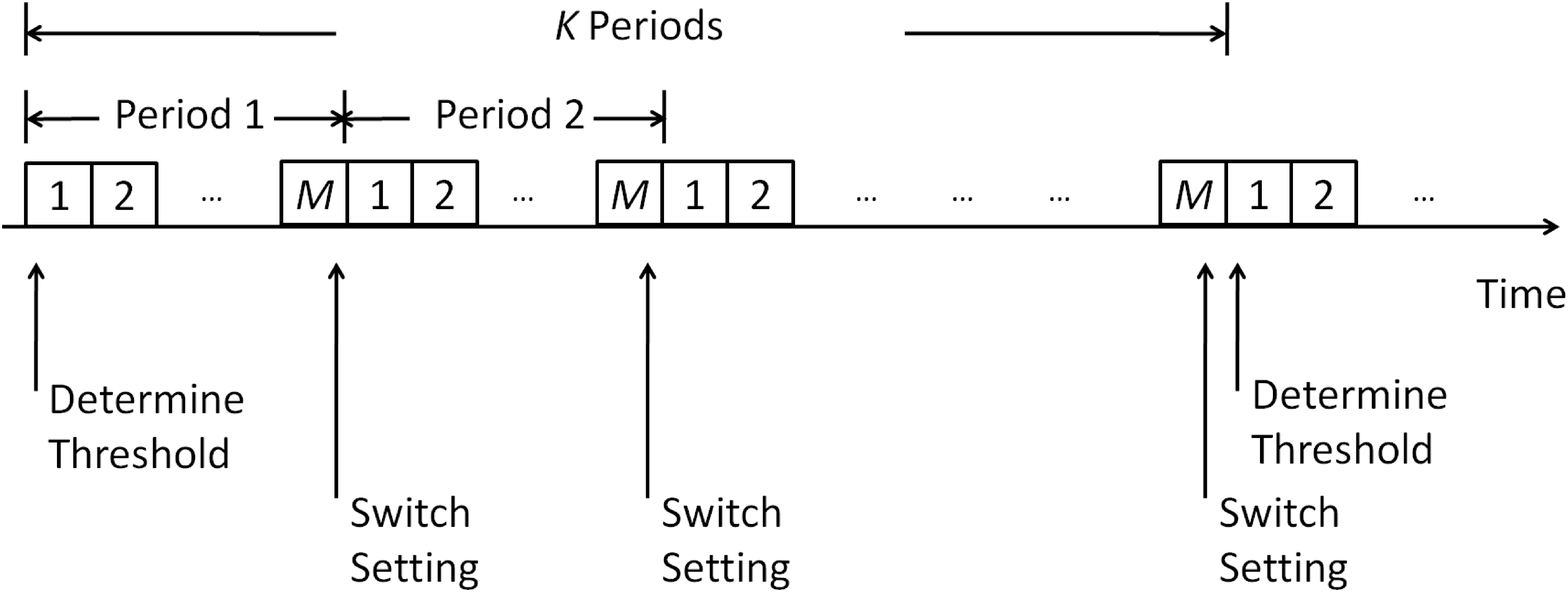}	
\caption{Time domain diagram of Prada-B algorithm}
\label{fig_prada_b}
\end{figure}

Since the FER in the previous setting adaptation period implies the wireless channel condition implicitly,  
we propose a FER-based algorithms.

The time domain diagram of this algorithm is shown in Figure \ref{fig_prada_b}.
A threshold adaptation period is $K$ setting adaptation periods.
At the end of a setting adaptation period,
the transmitter compares the FER to some predetermined threshold 
to decide which setting to used in next setting adaptation period.
At the beginning of a threshold adaptation period,
the transmitter decide the threshold used in the next $K$ setting adaptation periods.
The detail of the threshold determination is given in Section \ref{derivation}. 

The adaptation consists of two parts, 
one is the adaptive modulation and coding switching, 
and the other is the threshold adaptation. 
The former is for the short-term channel variation 
while the latter is for the long-term channel variation.

\subsubsection{Adaptive Modulation and Coding  Switching} \label{short_term}
The transmitter computes the FER in the previous setting adaptation period first, and then compare the FER to the threshold.
A threshold set consists of $2R$ values $\{T_{r, r-1}, T_{r, r+1}\}_{r=1}^R$, where $T_{1,0}=0$ and $T_{R, R+1}=1$. 
If the FER is lower than $T_{r, r-1}$ when using setting $s_r$, the transmitter switches to setting $s_{r-1}$; 
if the FER is higher than $T_{r, r+1}$ when using setting $s_r$, the transmitter switches to setting $s_{r+1}$; 
otherwise, the  transmitter uses the same setting as before. 
At the beginning, the system can start with any setting 
because it would switches to the optimal setting later.

The steps are summarized as follows:
\begin{itemize}
\item Step 1: Initiate the transmitter using setting $s_1$.
\item Step 2: In each setting adaptation period, record number of erroneous frames $l$, and the FER $=l/M$. 
\item Step 3:If the FER is lower than $T_{r, r-1}$, switch to setting $s_{r-1}$;
else if the FER is higher than $T_{r, r+1}$, switch to setting $s_{r+1}$;
else the setting remains the same. 
\end{itemize}

\subsubsection{Threshold Adaptation}\label{long_term}
Due to the large-scale fading of wireless channel, the transmitter changes the thresholds every $K$ setting adaptation periods (i.e., $KM$ frame transmissions). The thresholds are chosen to maximize the expected throughput of the next $KM$ frame transmissions. The transmitter changes the thresholds at the beginning of the $KM$ frame transmissions based on the channel state $w_i$ and the current setting $s_r$. The detail of computing the thresholds is given in Section \ref{derivation}

Here we summarize the steps as follows:
\begin{itemize}
\item Step 1: For each possible combination \{$s_rw_i$\}, find the corresponding threshold set. 
This can be done offline.
\item Step 2: At the beginning, use the threshold set corresponding to \{$s_1w_i$\} for the first $KM$ frame transmissions, where $w_i$ is the channel state at the beginning.
\item Step 3: After $K$ setting adaptation periods, the system changes the threshold set according to the first state \{$s_rw_i$\} of the next $K$ setting adaptation period.
\item Step 4: Repeat 3.
\end{itemize}

\section{Algorithm Analysis and Derivation} \label {derivation}

In this section, we describe the analysis of the PRADA algorithm.
Since PRADA-A is a simplified version of PRADA-B,
it only related to Section \ref{derive_throughput}.

\begin{table}
\center
\extrarowheight=2pt
\begin{tabular}{ll}
\hline
Parameter & Meaning \\
\hline
$R$ 				&Number of settings 	\\
$s_r$			& $r$-th setting \\
$k_r$			& Data rate of $s_r$ \\
$w_i$			& $i$-th channel state\\
\multirow{2}{*}{$M$}			& A setting adaptation period consists of $M$ frame \\ 					&transmissions\\
\multirow{2}{*}{$K$}			& A threshold adaptation period is $K$ setting adaptation \\ 						&periods\\
$E^{w_i}_{s_r}$ 	&FER when using $s_r$ in $w_i$ 	\\
$F_{i,k}^w(s_r,l)$	&$P[l \mbox{ frame errors}, M\mbox{-th state is }w_k|s_r, 1\mbox{st state is }w_i]$ \\
$T_{r,r\pm1}$	&Threshold to switch setting from $s_r$ to $s_{r\pm1}$ \\
\multirow{2}{*}{$P^w_{i,j}$}		&Probability of state change from $w_i$ to $w_j$ in a frame\\
				&transmission\\
$\Lambda_t$		&Induced Markov chain which models $w_i$ and $s_r$\\
$\lambda_{(i)}$	& $i$-th element in the state space of $\Lambda$\\
$P_{\lambda_{(i)},\lambda_{(j)}}$ & Transition probability of $\Lambda$\\
\multirow{2}{*}{$\xi_{\lambda_{(i)}}$}	&Expected throughput of the next $M$ frames when the\\ 
	&first state is $\lambda_{(i)}$\\
\hline
\end{tabular}
\caption{Parameters and their meanings}
\label{variable_table}
\end{table}

\subsection{Induced Markov Chain} \label{induced_mc}
\begin{figure}
\centering
\includegraphics[width=0.5\textwidth]{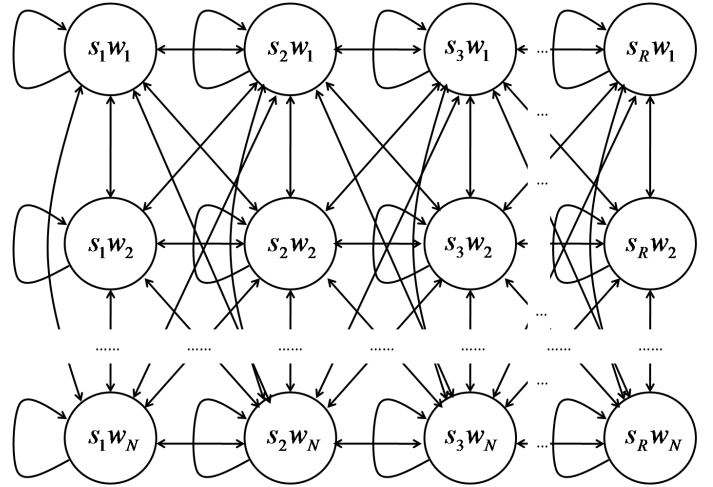}	
\caption{Induced Markov chain which models both channel variation and setting switching}
\label{fig_induced_mc}
\end{figure}

Let $\mathbf{P^w}$ denote the channel-state transition matrix, where $P^w_{i,j}$ is the probability that channel state changes from $i$ to $j$ in a frame transmission.
Let $F^w_{i,k}(s_r,l)$ be the probability that in a setting adaptation period, channel state ends at $w_k$ and $l$ frames are erroneous when using setting $s_r$ given that channel state starts from $w_i$. The derivation of $F^w_{i,k}(s_r,l)$ is given in Appendix \ref{derivation_of_F}. 

Let $\{\Lambda_t\}$ denote the induced Markov chain which models both setting switching and the varying wireless channel states. Each state represents the first state of 
a setting adaptation period. Denote the state space of $\{\Lambda_t\}$ by 
$S_\Lambda = \{s_rw_i| r \in \{1,2,\dots,R\}, i \in \{1, 2, \dots, N\}\} = \{\lambda_{(m)}\}$ where $\lambda_{(m)} = s_{(m)}w_{(m)}$ denotes the $m$-th element of the indexed state space.
Here we derive the transition probability matrix of $\Lambda_t$.

If the first state of this setting adaptation period is $s_rw_i$ and the first state of the next setting adaptation period is $s_{r-1}w_j$, it means:
\begin{enumerate}
\item the number of erroneous frames $l$ in this setting adaptation period satisfies that $l/M<T_{r,r-1}$
\item the channel state changes from some state to $w_j$ at the beginning of the next setting adaptation period.
\end{enumerate}
Hence, the transition probability can be derived as
\begin{displaymath}
P_{s_rw_i,s_{r-1}w_j}=\sum_{k=1}^{N}\sum_{l=0}^{\lfloor MT_{r,r-1}\rfloor}F^w_{i,k}(s_r,l)P^w_{k,j}.
\end{displaymath}
We can derive $P_{s_rw_i,s_rw_j}$ and $P_{s_rw_i,s_{r+1}w_j}$ by the same method.
Thus, the transition probabilities can be express as 
\begin{equation}
\begin{split}
&P_{s_rw_i,s_{r-1}w_j}=\sum_{k=1}^{N}\sum_{l=0}^{\lfloor MT_{r,r-1}\rfloor}F^w_{i,k}(s_r,l)P^w_{k,j}\\
&P_{s_rw_i,s_rw_j}=\sum_{k=1}^{N}\sum_{l=\lfloor MT_{r,r-1}\rfloor+1}^{\lfloor MT_{r,r+1}\rfloor}F^w_{i,k}(s_r,l)P^w_{k,j}\\
&P_{s_rw_i,s_{r+1}w_j}=\sum_{k=1}^{N}\sum_{l=\lfloor MT_{r,r+1}\rfloor+1}^{M}F^w_{i,k}(s_r,l)P^w_{k,j}\\
\end{split}
\end{equation}
As a result, we get the transition probability matrix of the induced Markov chain $\mathbf{P}$. The transition diagram are shown in Figure \ref{fig_induced_mc}.

\subsection{Expected Throughput} \label{derive_throughput}
Let $\xi_{s_rw_i}$ denote the expected throughput of $M$ frame transmissions when the first state is $s_rw_i$. Recall that $R$ settings $\{s_1, s_2, \dots, s_R\}$ are in the descending order of the FER. Each setting $s_r$ corresponds to a data rate $k_r$ which combines the rate of modulation and coding. In general,  $\{k_1, k_2, \dots, k_R\}$ are also in the descending order. For each frame, we define its expected throughput as $k_r(1-E^{w_i}_{s_r})$. Hence, the expected throughput for $M$ frames is
\begin{equation}
\xi_{s_rw_i} =k_r\sum_{k=1}^{N}\sum_{l=0}^{M}\frac{M-l}{M}F^w_{i,k}(s_r,l).
\label{eq_xi}
\end{equation}
Note that $F^w_{i,k}(s_r,l)$ contains the probability that the channel state ends at $w_k$ and the probability that $l$ frames are in error.

\subsection{Threshold}
The throughput of $KM$ frames is defined as
\begin{equation}
\xi_T=\frac{1}{K}\sum_{t=1}^{K}\xi_t
\end{equation}
where $\xi_t$ is the average throughput of $t$-th period. The expected throughput of $KM$ frames given $\Lambda_1$ is

\begin{equation}
\begin{split}
E[\xi_T|\Lambda_1] &= \frac{1}{K}\sum_{t=1}^KE[\xi_t|\Lambda_1]\\
&= \frac{1}{K} \sum_{t=1}^K \sum_{m=1}^{RN}P[\Lambda_t= \lambda_{(m)}]\xi_{\lambda_{(m)}}\\
&=\frac{1}{K} \sum_{t=1}^K\sum_{m=1}^{RN}P_{\Lambda_1, \lambda_{(m)}}^{(t-1)}\xi_{\lambda_{(m)}}
\end{split}
\end{equation}
where $P_{\lambda_{(i)}, \lambda_{(j)}}^{(n)} = P[\Lambda_{m+n}=\lambda_{(j)}|\Lambda_{m}=\lambda_{(i)}]$ is the  $n$-step transition probability and $\xi_{\lambda_{(m)}}$ is given in Equation (\ref{eq_xi}). 
Recall that $\mathbf{P}^{(n)}=\mathbf{P}^n$ only depends on the threshold set, 
as that described in Section \ref{induced_mc}. Thus,
we can choose the threshold set such that $E[\xi_T|\Lambda_1]$ is maximized for every 
possible $\Lambda_1$.

An algorithm used to find the threshold set is described as following.
For each possible $\Lambda_1$
\begin{enumerate}
\item Generate a threshold set $\{T_{r,r-1}, T_{r,r+1}\}_{r=1}^R$ as the current threshold set randomly.
\item Change one value in the current threshold set by 1 or $-1$ to get a new threshold set. Do this for each value except for $T_{1,0}$ and $T_{R, R+1}$ and get $2R-2$ new
threshold sets.
\item Compute the expect throughput for $2R-2$ new threshold sets. Choose the 
threshold set that maximizes the expect throughput to be the current threshold set. 
\item Repeat step 2-3 until all $2R-2$ new threshold sets are not better than the
current threshold set. Then output the current threshold set.
\item Do step 1-4 several times, choose the best threshold set among the output threshold sets.
\end{enumerate}


\section{Numerical Result}

\subsection{Performance of Different Average SNR}

\begin{table}
\center
\extrarowheight=1pt
\begin{tabular}{cccc}
\hline
State index & $P^w_{i,i-1}$ & $P^w_{i,i}$ & $P^w_{i,i+1}$\\
\hline
1 	&0.0000 	&0.9893		&0.0107\\
2 	&0.0155 	&0.9736 	&0.0109 \\
3 	&0.0166 	&0.9737 	&0.0097 \\
4 	&0.0177		&0.9737 	&0.0086 \\
5 	&0.0187 	&0.9737 	&0.0076 \\
6 	&0.0198 	&0.9736 	&0.0066 \\
7 	&0.0159 	&0.9841		&0.0000 \\
\hline
\end{tabular}
\caption{Transition probabilities of the FSMC.}
\label{table_channel_model}
\end{table}

\begin{table}
\center
\extrarowheight=1pt
\begin{tabular}{cc}
\hline
Boundary & Value (dB)\\
\hline
$\Gamma_1$ 	&$-\infty$ \\
$\Gamma_2$ 	&$2.0499$ \\
$\Gamma_3$ 	&$4.0232$ \\
$\Gamma_4$ 	&$5.6514$ \\
$\Gamma_5$ 	&$7.0454$ \\
$\Gamma_6$ 	&$8.2726$ \\
$\Gamma_7$ 	&$9.3777$ \\
$\Gamma_8$ 	&$\infty$ \\
\hline
\end{tabular}
\caption{SNR threshold values of the FSMC}
\label{table_bound}
\end{table}

\begin{table}
\center
\extrarowheight=1pt
\begin{tabular}{ccccc}
\hline
Setting	&Modulation	&Code rate	&Frame size	&Data bits/Frame ($k_r$)\\
\hline
$s_1$	&16-QAM	&2/3		&8192	&5461\\
$s_2$	&16-QAM	&1/2		&8192	&4096\\
$s_3$	&QPSK		&2/3		&4096	&2731\\
$s_4$	&QPSK		&1/2		&4096	&2048\\
$s_5$	&BPSK		&1/2		&2048	&1024\\
\hline
\end{tabular}
\caption{The settings used in the simulation. The frame size corresponds to 2048 symbols.}
\label{table_amc_scheme}
\end{table}

\begin{table}
\center
\extrarowheight=1pt
\begin{threeparttable}
\begin{tabular}{cccccc}
\hline
State 	&$s_1$	&$s_2$	&$s_3$	&$s_4$	& $s_5$\\
\hline
1 	&1.0000	&1.0000	&1.0000		&1.0000		&1.0000	\\
2 	&1.0000	&1.0000	&0.7741		&0.5406 	&0.3662 \\
3 	&1.0000	&0.9387	&0.0900		&0.0229 	&0.0128 \\
4 	&0.9193	&0.3752	&0.0020		&0.0000 	&0.0000 \\
5 	&0.3807	&0.0411	&0.0000 	&0.0000 	&0.0000	\\
6 	&0.0609	&0.0026	&0.0000 	&0.0000 	&0.0000	\\
7 	&0.0045	&0.0000	&0.0000 	&0.0000		&0.0000	 \\
\hline
\end{tabular}
\begin{tablenotes}
\item[*] The zero values are actually very small numbers thus neglected. 
\end{tablenotes}
\end{threeparttable}
\caption{FER of each setting in different channel states}
\label{table_fer_in_each_state}
\end{table}

In this channel model, the doppler frequency is set to 4 Hz and the 
average SNR is set to 2 dB.
The FSMC used to simulate the wireless channel is shown in
Table \ref{table_channel_model}. 
The SNR boundary values in the FSMC are computed according to Section \ref{sec_channel_model}, listed in Table \ref{table_bound},
where state $w_i$ is between $\Gamma_i$ and $\Gamma_{i+1}$. 
When channel is in $w_1$, the transmitter does not transmit signals.

In the simulation, we consider five settings.
Table \ref{table_amc_scheme} shows the details of the five settings.
Note that the frame durations of all settings are equal.
The simulated FER values are listed in Table \ref{table_fer_in_each_state}.

\begin{table}
\center
\extrarowheight=1pt
\begin{tabular}{lr}
\hline
\multirow{2}{*}{Scenario} & Average throughput \\
	& (Data bits/Frame)\\
\hline
PRADA-A 		&1829.7\\
PRADA-B		&1873.9\\
16-QAM (2/3)	&527.6\\
16-QAM (1/2)	&920.8\\
QPSK (2/3)		&1724.8\\
QPSK (1/2)		&1544.4\\
BPSK (1/2)		&854.2\\
\hline
\end{tabular}
\caption{The average throughputs of different scenarios.}
\label{table_throughput}
\end{table}

To evaluate the performance of the PRADA algorithm, 
we simulate the adaptation case (PRADA-A, $M=120$; PRADA-B, $K=4$, $M=30$) and 
the no adaptation cases ($s_1$, $s_2$, $s_3$, $s_4$, and $s_5$).
We set the PRADA-A and the PRADA-B to have the equal frequency to obtain
channel state information. That is, the feedback rates of both PRADA-A and PRADA-B are the same (per 120 frames).
For fair comparisons, all simulation scenarios experience 
the same channel variations.
In the simulation, 5,000,000 frames are transmitted, 
the average throughputs over time are shown in Table \ref{table_throughput}.

\begin{figure}
\centering
\includegraphics[width=0.5\textwidth]{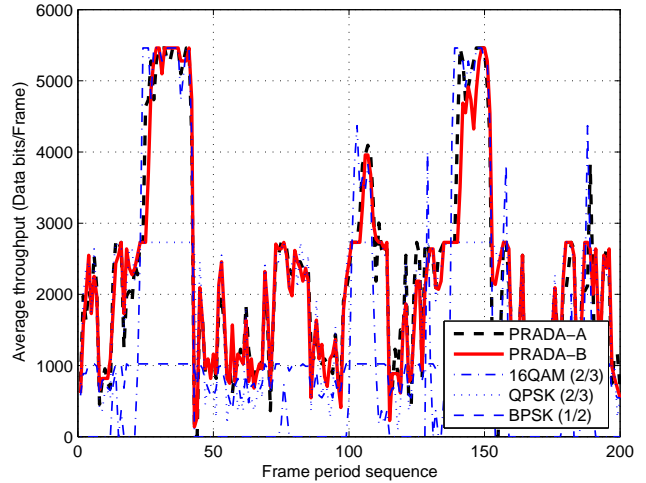}	
\caption{Throughput over time. Each point is the average throughput of 30 frames.}
\label{fig_time_result}
\end{figure}

The result shows that the performance of the PRADA is better
than that of the fixed modulation. 
The throughput over time is shown in Figure \ref{fig_time_result}.
We observe that while the performance of the fixed modulation saturates, PRADA algorithm can switch to the high-rate setting.
When the channel is in bad condition, the PRADA algorithm can switch to the
low-rate setting to maintain an acceptable performance.

\begin{figure}
\centering
\includegraphics[width=0.5\textwidth]{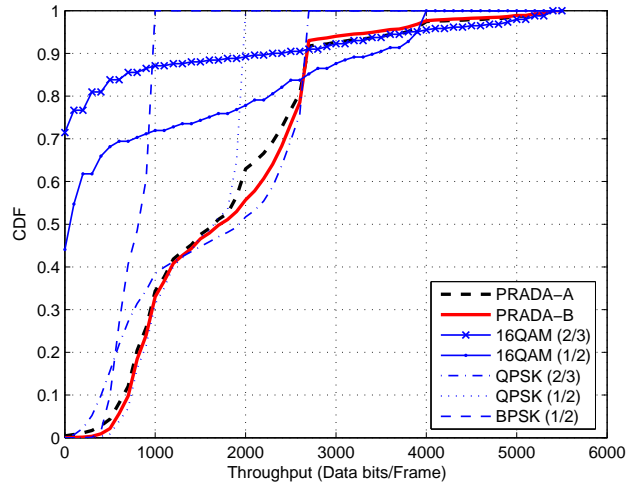}	
\caption{CDF comparison of all cases}
\label{fig_cdf}
\end{figure}

Figure \ref{fig_cdf} shows the throughput cumulative distribution
function (CDF) comparison of all cases. The CDF curves of the PRADA
is better than other curves. At the low throughput, the curve of the PRADA
is close to the curve of the QPSK (1/2); at the median throughput,
the curve of the PRADA is close to the curve of QPSK (2/3);
at the high performance, the curve of the PRADA is close to the curve of
16QAM (2/3). This implies the adaptation of the PRADA algorithm.

\begin{figure}
\centering
\includegraphics[width=0.5\textwidth]{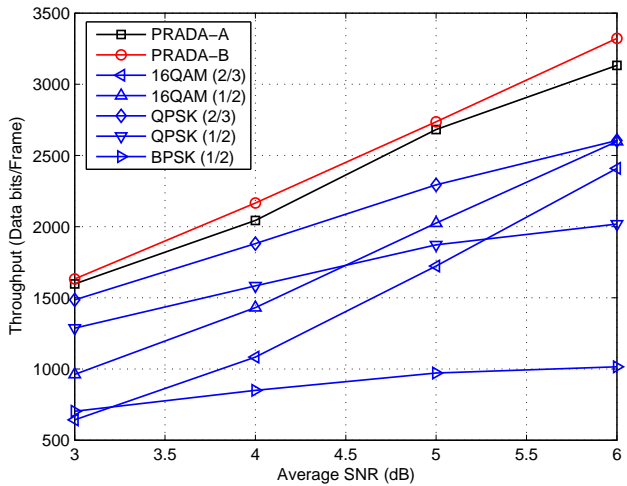}	
\caption{Throughputs of different average SNR}
\label{fig_snr}
\end{figure}

\begin{figure}
\centering
\includegraphics[width=0.5\textwidth]{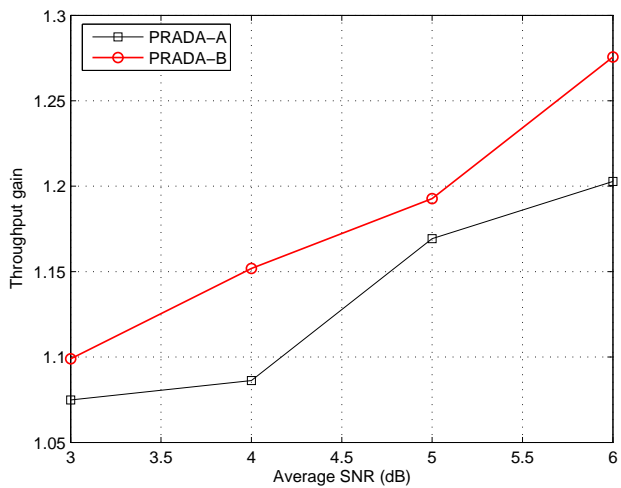}	
\caption{Throughput gain of PRADA algorithm}
\label{fig_throughput_gain}
\end{figure}

Throughputs of different average SNR are plotted in Figure \ref{fig_snr}.
To get the gain of PRADA algorithm, we compute the ratio of the throughput of PRADA to the highest throughput among the five fixed settings.
Figure \ref{fig_throughput_gain} shows the throughput gain of PRADA-A and PRADA-B.

\subsection{Performance of Different Doppler Frequencies}

\begin{table}
\center
\extrarowheight=1pt
\begin{threeparttable}
\begin{tabular}{ccc}
\hline
State index & $P^w_{i,i-1}$ &  $P^w_{i,i+1}$\\
\hline
1 	&$0$	&$0.00268f$\\
2 	&$0.00386f$ 	 	&$0.00271f$ \\
3 	&$0.00416f$	 	&$0.00242f$ \\
4 	&$0.00442f$		&$0.00216f$ \\
5 	&$0.00468f$ 		&$0.00190f$ \\
6 	&$0.00493f$ 	 	&$0.00164f$ \\
7 	&$0.00396f$ 		&$0$ \\
\hline
\end{tabular}
\begin{tablenotes}
\item[*] $f$ is the doppler frequency in Hz. 
\item[*] $P^w_{i,i} = 1- (P^w_{i,i-1}+P^w_{i,i+1})$
\end{tablenotes}
\end{threeparttable}
\caption{Transition probabilities of the FSMC.}
\label{table_channel_doppler}
\end{table}

In this simulation, the average channel SNR is set to 2dB,
and the doppler frequency is from 2 Hz to 20 Hz.
This simulation compares the following three algorithms.
\begin {enumerate}
\item PRADA-A ($M=120$)
\item PRADA-B ($M=30$, $K = 4$)
\item A compared AMC algorithm that chooses the setting such that the throughput of the first frame is maximized. The setting is switched every 120 frames.
\end{enumerate}
Note that three algorithms above have the same rate of the CSI feedback.
The settings used in the simulation is listed in Table \ref{table_amc_scheme}.
The FSMC channel boundaries are listed in Table \ref{table_bound}
and the transition probabilities are list in Table\ref{table_channel_doppler}.

Figure \ref{fig_throughput_gain} shows the results of different doppler frequencies. PRADA-B is always better than other two algorithm.
When the doppler frequency becomes high, the performances of
three algorithms decrease.
However, PRADA-A and PRADA-B at least maintain a performance that 
the best fixed setting achieves.
At high doppler frequency, the compared algorithm 
may be worse than the fixed setting.

The performances of these three algorithms when the doppler frequency
varies with time are shown in Table \ref{table_doppler_avg}.
The doppler frequency changes every $KM$ frames
and it varies from 2 Hz, 4Hz, ... to 20 Hz randomly.

\begin{figure}
\centering
\includegraphics[width=0.5\textwidth]{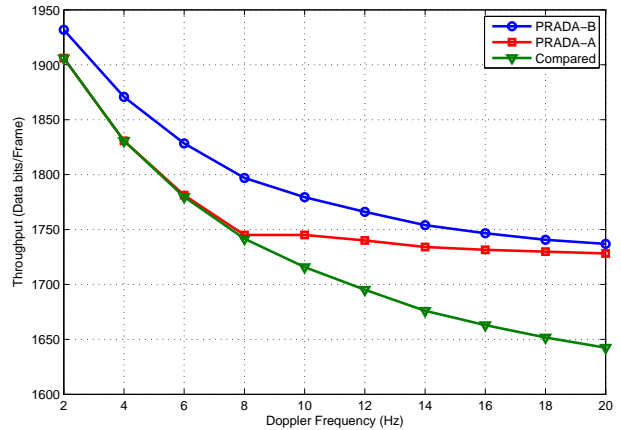}	
\caption{Throughput of different doppler frequencies }
\label{fig_performance_doppler}
\end{figure}

\begin{table}
\center
\extrarowheight=1pt
\begin{tabular}{cc}
\hline
Algorithm	& Throughput (data bits/frame)\\
\hline
PRADA-A 		&1768\\
PRADA-B		&1792\\
Compared Algorithm	&1731\\
\hline
\end{tabular}
\caption{The throughput with doppler frequency variation}
\label{table_doppler_avg}
\end{table}

\subsection{Analytical Results}
\begin{figure}
\centering
\includegraphics[width=0.5\textwidth]{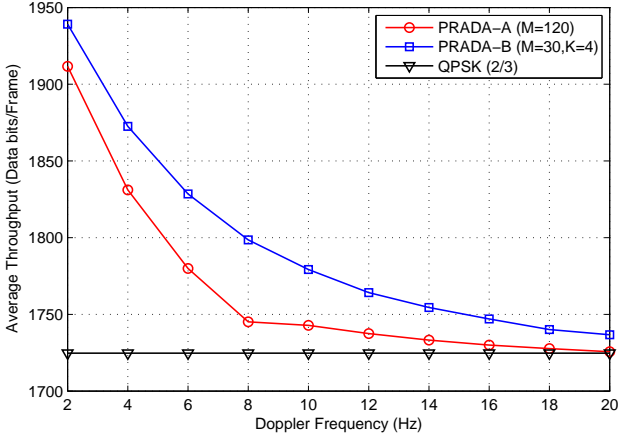}	
\caption{Analytical results of throughput with doppler frequency variation.}
\label{fig_analy_result}
\end{figure}

\begin{table}
\center
\extrarowheight=1pt
\begin{threeparttable}
\begin{tabular}{ccc}
\hline
Doppler frequency (Hz)& PRADA-A & PRADA-B\\
\hline
2	&1911.65	&1939.18\\
4	&1831.16	&1872.55\\
6	&1779.94	&1828.52\\
8	&1745.16	&1798.56\\
10	&1742.85	&1779.25\\
12	&1737.48	&1764.22\\
14	&1733.25	&1754.44\\
16	&1730.02	&1747.02\\
18	&1727.65	&1740.13\\
20	&1725.66	&1736.69\\
\hline
\end{tabular}
\begin{tablenotes}
\item[*] Throughput unit: data bits/frame
\end{tablenotes}
\end{threeparttable}
\caption{Analytical results of throughput with doppler frequency variation.}
\label{table_analy_result}
\end{table}

Here we provide the analytical results for comparison.
Figure \ref{fig_analy_result} shows the analytical value of PRADA performance.
It is observed that these curve match that in figure \ref{fig_performance_doppler}.
Table \ref{table_analy_result} also show the analytical value of PRADA performance.
The throughputs when doppler frequency is 4Hz match that in Table \ref{table_throughput}.


\section{Conclusion}
In this paper, the problem of the feedback frequency reduction has been formally addressed and investigated. A close form expression of expectation FER over arbitrary $M$ frames is provided. On the foundation of the derivation, we were able to propose two adaptive algorithms, PRADA-A and PRADA-B with different degrees of feedback reduction. It was shown that the first algorithm proposed achieved significant performance improvement over fixed modulation. The second algorithm further reduced the frequency of feedback, moreover, introduced a novel adaptive transmission technique combing the feedback CSI and FER. The algorithms introduced in this work allow the system designer to flexibly switch the adaptation period according to the application under consideration.

\appendix
\subsection{Derivation of $\{F^w_{i,k}(s_r, l)\}$}\label{derivation_of_F}
First we define a random variable 
\[
X_j=\left\{
\begin{array}{cl}0, & \mbox{successful transmission}\\
1, & \mbox{erroneous transmission} \end{array} \right.
\]
Recall that $E^{w_i}_{s_r}$ is the FER when using the setting $s_r$ in the channel state $w_i$. Given the used setting $s_r$ and the channel state $w_i$, the probability generating function (PGF) of $X_j|s_r,w_i$ is
\begin{equation}
\begin{split}
\phi_{X_j|s_r,w_i}(\omega) &= \sum_{x=0, 1}\omega^xP[X_j=x]\\
&= (1-E^{w_i}_{s_r})+\omega E^{w_i}_{s_r}\\
&\triangleq \psi_{w_i}^{s_r}(\omega)
\end{split}
\end{equation}
Denote the channel states in $M$ frame transmissions as a random vector
$\mathbf{w}=[w_{i_1} w_{i_2} \dots w_{i_M}]$. The conditional PGF
of the total number of erroneous frame in $M$ frame transmissions is
\begin{equation}
\begin{split}
&\phi_{X_1+X_2+\dots+X_M|s_r,\mathbf{w}}(\omega)\\
&= \phi_{X_j|s_r,w_{i_1}}(\omega)\phi_{X_j|s_r,w_{i_2}}(\omega)\dots\phi_{X_j|s_r,w_{i_M}}(\omega)\\
&= \psi_{w_{i_1}}^{s_r}(\omega) \psi_{w_{i_2}}^{s_r}(\omega) \dots \psi_{w_{i_M}}^{s_r}(\omega)
\end{split}
\end{equation}
We need to take expectation of the conditional PGF with respect to the probability of every possible $\mathbf{w}$.
This can be done by combining transition matrix and 
$\{\psi_{s_r,w_i}(\omega)\}$ as
\begin{equation}
\begin{split}
&G_{s_r}(\omega)\\
&= \mathbf{P^w}\psi_{s_r}(\omega)\\
&=\left[
\begin{array}{cccc}
P^w_{1,1}\psi^{s_r}_{w_1}(\omega) & P^w_{1,2}\psi^{s_r}_{w_2}(\omega) & \cdots & 0\\
P^w_{2,1}\psi^{s_r}_{w_1}(\omega) & P^w_{2,2}\psi^{s_r}_{w_2}(\omega) & \cdots & 0\\
\vdots & \vdots & \vdots & \vdots\\
0 & 0 & \cdots & P^w_{N,N}\psi^{s_r}_{w_N}(\omega)
\end{array}
\right]
\end{split}
\end{equation}
where
\begin{equation}
\psi_{s_r}(\omega)\triangleq
\left[
\begin{array}{cccc}
\psi^{s_r}_{w_1}(\omega) & 0 & \cdots & 0\\
0 &\psi^{s_r}_{w_2}(\omega) & \cdots & 0\\
\vdots & \vdots & \vdots & \vdots\\
0 & 0 & \cdots & \psi^{s_r}_{w_N}(\omega)
\end{array}
\right]
\end{equation}
We can find the PGF of erroneous frame number in a setting adaptation period by matrix multiplication.
That is,
\begin{equation}
\begin{split}
H_{s_r}(\omega)&\triangleq\psi_{s_r}(\omega)G_{s_r}(\omega)^{M-1}\\
&=\left[
\begin{array}{cccc}
H_{1,1}^{s_r}(\omega) & H_{1,2}^{s_r}(\omega) & \cdots & H_{1,N}^{s_r}(\omega)\\
H_{2,1}^{s_r}(\omega) & H_{2,2}^{s_r}(\omega) & \cdots &H_{2,N}^{s_r}(\omega)\\
\vdots & \vdots & \vdots & \vdots\\
H_{N,1}^{s_r}(\omega) &H_{N,2}^{s_r}(\omega) & \cdots & H_{N,N}^{s_r}(\omega)
\end{array}
\right]
\end{split}
\end{equation}
Note that $H_{i,j}^{s_r}(\omega)$ is simply the PGF of $F^{w}_{i,j}(s_r,l)$.
That is,
\begin{equation}
H_{i,j}^{s_r}(\omega) = \sum_{l=0}^MF^{w}_{i,j}(s_r,l)\omega^l.
\end{equation}
One can find \{$F^{w}_{i,j}(s_r,l)$\} easily by $H_{i,j}^{s_r}(\omega) $


\end{document}